\documentclass[aps,prx,twocolumn,superscriptaddress]{revtex4}

\usepackage{graphicx}
\usepackage{bm}        % for math
\usepackage{amsmath}
\usepackage{amssymb}   % for math
\usepackage{array}
\usepackage[bookmarks=false,linkcolor=blue,urlcolor=blue,colorlinks,citecolor=blue]{hyperref}
\def\nn{\nonumber}
\newcommand{\beq}{\begin{eqnarray}}
\newcommand{\eeq}{\end{eqnarray}}
\newcommand{\mc}[1]{\mathcal{#1}}

\makeatletter

\begin{document}

\title{Theory of anomalous magnetotransport from mass anisotropy}

\author{Liujun Zou}

\affiliation{Department of Physics, Harvard University, Cambridge, MA 02138}
\affiliation{Department of Physics, Massachusetts Institute of Technology, Cambridge, MA 02139}

\author{Samuel Lederer}

\affiliation{Department of Physics, Massachusetts Institute of Technology, Cambridge, MA 02139}

\author{T. Senthil}
\affiliation{Department of Physics, Massachusetts Institute of Technology, Cambridge, MA 02139}

\date{\today}

\begin{abstract}
 In underdoped YBa$_2$Cu$_3$O$_{6+x}$, there is evidence of a small Fermi surface pocket subject to substantial mass enhancement in the doping regime $ 0.12<p<0.16$. This mass enhancement may vary substantially over the Fermi surface, due to ``hot spot" or other relevant physics. We therefore examine the magnetotransport of an electron-like Fermi pocket with large effective mass anisotropy. Within the relaxation time approximation, we show that even for a pocket with a fixed shape, the magnitude and sign of the Hall effect may change as the mass anisotropy changes (except at very large, likely inaccessible magnetic fields). We discuss implications for recent Hall measurements in near optimally doped cuprates in high fields. In addition we identify a novel intermediate asymptotic regime of magnetic field, characterized by B-linear magnetoresistance. Similar phenomena should occur in a variety of other experimental systems with anisotropic mass enhancement.% in which carriers propagate uninterrupted on the ``light" parts of the Fermi surface but scatter rapidly on the``heavy" parts.K

\end{abstract}
\maketitle

\section{Introduction}

Recent high magnetic field experiments\cite{Sebastian2015} have shed new light on the underdoped regime of several cuprate high temperature superconductors (particularly YBa$_2$Cu$_3$O$_{6+x}$ or YBCO), revealing a field-induced metallic state at low temperature that exhibits quantum oscillations, a hallmark of Fermi liquid behavior.  A precise theoretical description of this phase (or phases) could provide a valuable framework by which to understand, among other phenomena, the pseudogap regime that prevails at higher temperature. However, even relatively basic questions about the phenomenology of the field-induced metallic regime, such as the number of phases present and the symmetries they break, remain unsettled.

In YBCO, one unsettled question concerns the extent of charge density wave (CDW) order within the metallic region of the phase diagram. Recent high field measurements\cite{Badoux2016} show a substantial variation of the Hall coefficient over doping levels $p$ between $0.16$ and $0.205$, consistent with the scenario of a quantum critical point (QCP) near $p=0.19$. Furthermore, the Hall coefficient is positive, opposite in sign to that observed at lower dopings.

How should we think about this sign change of the Hall coefficient?
Since the negative Hall coefficient for  $p<0.16$ is associated with CDW order\cite{leboeuf2011}, the authors of Ref. \cite{Badoux2016} suggest that CDW order terminates at or below $p=0.16$. In this understanding, the region between $0.16<p<0.19$ contains, at high fields, a distinct metallic phase featuring small hole pockets. A candidate state with such hole pockets is a spin density wave metal--however to date there is no indication of long range spin density wave order at these doping levels in YBCO in {\em zero} magnetic field. Whether such order is induced by the magnetic field is not currently known, and is a good target for future experimental work. Another class of proposals \cite{Yang2006,Stanescu2006,Senthil2008,Mei2012,Chowdhury2014} posit an interesting metallic state which does not break any symmetries, but nevertheless has small hole pockets violating the standard Luttinger theorem. Such a state necessarily has fractionalized excitations in addition to the Fermi pockets, and hence is known as a Fractionalized Fermi Liquid\cite{Senthil2003}. Clear evidence in support of a Fractionalized Fermi Liquid in the cuprates would be a tremendously exciting development, and is again a fascinating target for future experimental work.

However, some caution is warranted in the interpretation of the results on the Hall coefficient. One factor is the elevated temperature ($T\sim50 K$) of the measurements, which makes them difficult to compare to measurements in the regime of quantum oscillations at lower doping and $T\sim 4 K$. Another factor is that the conventional interpretation of the Hall coefficient, as a measure of the number and sign of charge carriers, may fail near the putative QCP. A breakdown of Fermi liquid and/or Boltzmann transport theory would naturally invalidate this conventional interpretation.
However, even if Fermi liquid and Boltzmann transport theory are valid,  it is not clear that the conventional interpretation is necessarily correct, as we demonstrate in this paper. Our result, along with previous work\cite{Harrison2015,Robinson2015}, raises the possibility that the small pocket that exists at lower doping persists all the way to $p = 0.19$ (presumably along with the CDW order), but nevertheless has a sign change of its measured Hall effect. It remains to be seen if this is what actually happens in YBCO, or if another state, such as those discussed in the previous paragraph, is realized.

We study a Fermi liquid metal in a situation where the quasiparticle effective mass varies strongly around the Fermi surface. In the context of the underdoped cuprates, precisely such a highly anisotropic effective mass was proposed\cite{Senthil2014} by one of us to account for seemingly conflicting measurements of the effective mass in underdoped cuprates. Here we study the magnetotransport properties of such a Fermi liquid metal within the standard Boltzmann framework.  To be concrete, we will treat a Fermi surface similar to the model proposed by Harrison and Sebastian\cite{Harrison2011}, in which a diamond-shaped electron pocket and CDW order are both present.  Following \cite{Senthil2014}, we will assume that there are `heavy' portions near the corner of the Fermi  pocket, and `light' portions near the zone diagonal.
  We have two principal results, which are not limited to the specific form of the model chosen. The first is that there is generically a change in sign of the Hall coefficient  as the ratio of the heavy to the light quasiparticle masses is increased at fixed magnetic field. The second is that when this ratio is large there exists a parametrically broad regime of magnetic fields, distinct from the familiar weak- and strong-field limits, in which simple Drude-like formulas badly mischaracterize the system. In this regime, the Hall number bears no systematic relation to the number of carriers or their charge, and the magnetoresistance is linear in the field, with a coefficient independent of disorder strength and of the effective mass.

 In Ref. \cite{Senthil2014} it was further proposed that this ratio of the heavy and light effective masses diverges as a putative quantum critical point around $ p = 0.19$ is approached. Within this proposal it follows from our results that there will generically be a sign change of the Hall effect as the doping is increased towards $0.19$ even if the CDW order persists, without any fundamental change in the Fermi surface topology.

Our results should be of broader interest in the theory of metals (apart from just the cuprates). Large variations of the quasiparticle effective mass may simply occur from band structure effects (such as  proximity to a van Hove singularity\cite{Norman2010,Lin2010}), but also from fluctuations which renormalize the band structure. An example is in metals proximate to a density wave instability. There, the soft density wave fluctuations will couple strongly to fermions at ``hot spots" where the ordering wave vector nests the Fermi surface, leading to an enhanced effective mass near the hot spots but little effect elsewhere.  Other examples are  heavy fermi liquids in rare earth alloys. The strong enhancement of quasiparticle effective mass that characterizes these metals likely occurs in some portions of the Fermi surface but not in others, thereby leading to strong variations of the effective mass around the Fermi surface\cite{Ghaemi2008}. Our results, for instance the regime of linear magnetoresistance, are pertinent to all such metals.

 The remainder of this paper is organized as follows: we introduce the model Fermi surface in section \ref{sec:model}; we then review the weak- and strong-field regimes in section \ref{sec:conv}, and the novel intermediate asymptotic regime in section \ref{sec:unc}; we close with a discussion of the implications of our results for the interpretation of experiments in the underdoped cuprates.

\section{Model of an anisotropic Fermi pocket}
\label{sec:model}

\begin{figure}
\centering
\includegraphics[clip=true, trim= 100 200 100 200, width=0.8\columnwidth]{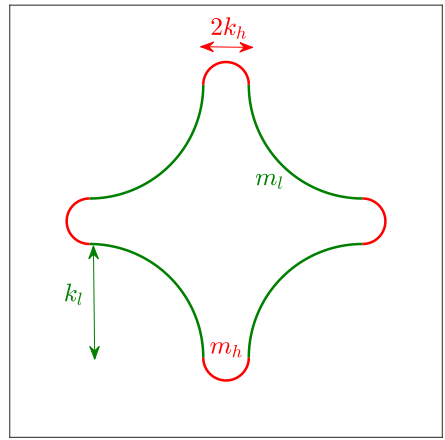}
\caption{The simple model for the Fermi surface pocket used in these notes. It consists of several circular arcs. The light segments in green have radius $k_l$ and Fermi velocity $v_l$, while the heavy segments in red have radius $k_h$ and Fermi velocity $v_h$.
}
\label{fig:model}
\end{figure}
Fig. \ref{fig:model} shows a simplified model of a 2D electron pocket with a diamond-like shape similar to that proposed by Harrison and Sebastian. It consists of circular arcs of radius $k_h$ and $k_l$ with Fermi velocities  $v_h$ and $v_l$ respectively\footnote{The piecewise definition here cannot arise from a smooth dispersion relation. This entails modest singularities in magnetic field dependence as $B\to 0$, which are not of concern here. For the numerical calculations, we round the jump in Fermi velocity between ``light" and ``heavy" segments.}, the subscripts $h$ and $l$ refer to ``heavy" and ``light".
For a given magnetic field $B$, the cyclotron orbits  have angular velocities $\omega_{h,l}\equiv eB\hbar v_{h,l}/k_{h,l}$, on the heavy and light segments, and the cyclotron period is
\begin{equation}
\frac{2\pi}{\omega_c}=\frac{4\pi}{\omega_h}+\frac{2\pi}{\omega_l}=\frac{2\pi \hbar}{eB}\left(\frac{2k_h}{v_h}+\frac{k_l}{v_k}\right)
\end{equation}
If we introduce the effective masses $m_{h,l}\equiv eB/\omega_{h,l}$, then the cyclotron effective mass is $m_c \equiv eB/\omega_c=2m_h+m_l$.

While this model is artificial, the conclusions we glean from it rest on only a few essential assumptions. The first is that the Fermi surface is an electron pocket consisting primarily of segments with hole-like curvature, an assumption common to most proposals\cite{Harrison2011,Yao2011} to explain high-field magnetotransport in YBCO. The second assumption is that the segments of electron-like curvature connecting the hole-like segments are especially subject to mass enhancement as the CDW transition is approached. This relies on the CDW terminating in a QCP (or weakly first order transition), and was previously argued \cite{Senthil2014} to account for seemingly conflicting measurements of the effective mass in underdoped cuprates. We will be interested in the behavior of the model on approach to a QCP at which $m_h/m_l$ diverges\footnote{$k_h$ would also vanish at the transition, but this effect changes none of our qualitative conclusions, so we fix $k_h=0.2 k_l$.}, and we will simplify the discussion by keeping $m_l$ fixed to equal the electron mass $m_e$.

We employ the relaxation time approximation to Boltzmann transport. For simplicity we work at zero temperature, with an isotropic relaxation time $\tau=0.2 ps$\cite{Ramshaw2012} \footnote{A finite value of $\tau$ at zero temperature derives from disorder scattering. For a fixed density of point-like scatterers, and assuming the coupling of quasiparticles to disorder is not singularly renormalized, $\tau$ will be proportional to the inverse density of states, and therefore tend to vanish as $m_c$ diverges. We neglect this effect here, but it may be important for the interpretation of transport experiments on approach to various QCPs.}. To compute the DC conductivity we use Chambers' formula \cite{Ashcroft1976}, which is valid for arbitrary magnetic field:
\begin{equation}
 \label{eq:Chambers0T}
\sigma_{ij}=\frac{e^2}{2\pi^2\hbar}\oint \frac{dk}{\left|\mathbf{v}(\mathbf{k})\right|} v_j(\mathbf{k})\int_0^\infty dt\ v_i(\mathbf{q}(t))e^{-t/\tau}
\end{equation}
Here the $\mathbf{k}$ integral is over the Fermi surface, $\mathbf{q}(t=0)=\mathbf{k}$, and the time evolution of $\mathbf{q}$ is given by the Lorentz force law $\hbar \dot{\mathbf{q}}=-e\mathbf{v}(\mathbf{q})\times \mathbf{B}=-eB\mathbf{v}(\mathbf{q})\times \hat z$.
\section{``Familiar" regimes}
\label{sec:conv}
\subsection{Weak field: $\omega_l\tau,\omega_h\tau\ll 1$}

\begin{figure}
\includegraphics[width=0.95\columnwidth]{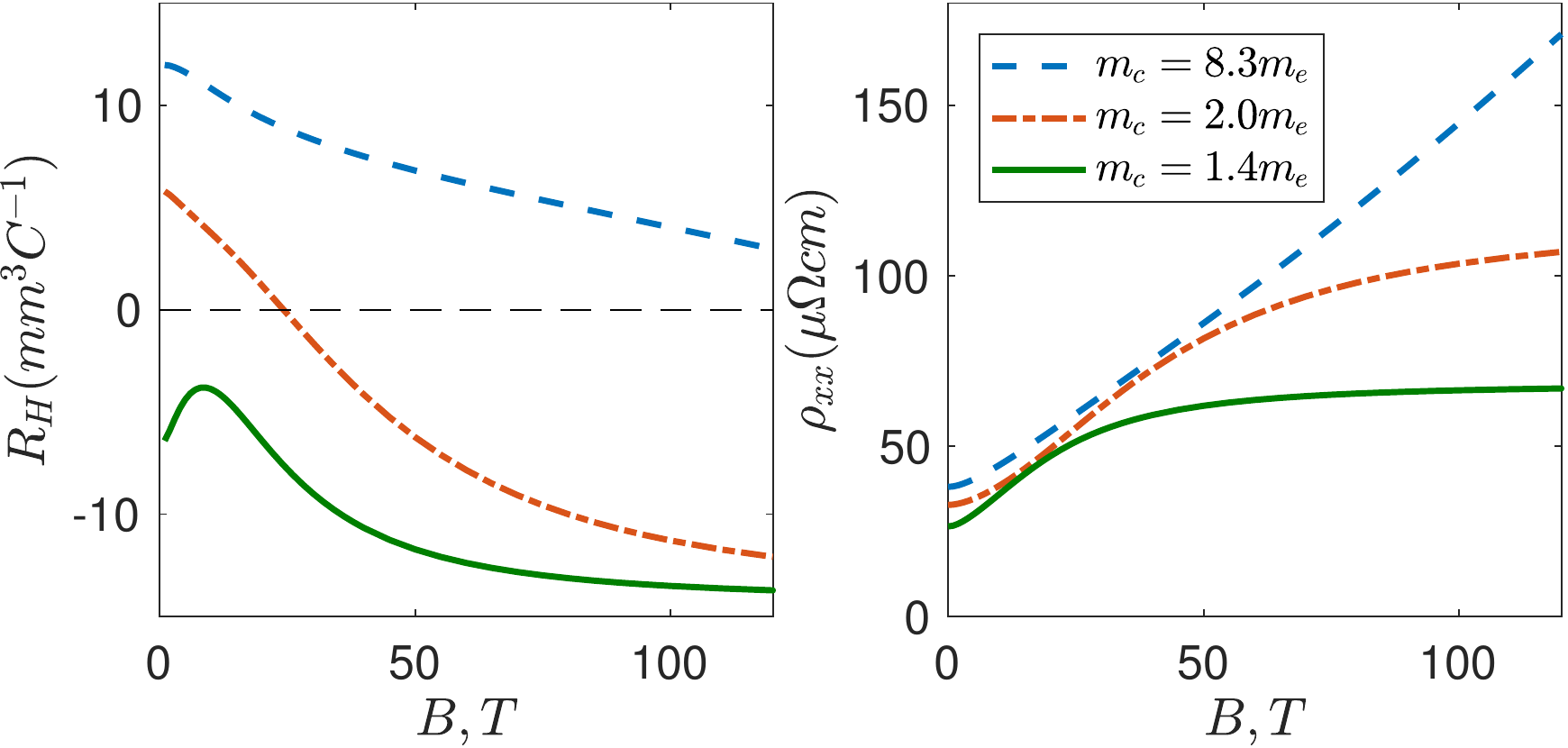}
\caption{Magnetic field dependence of the Hall coefficient (left) and longitudinal resistivity (right) for the model with fixed $m_l=m_e$ and various $m_h$. In the calculations, $k_h=0.2k_l$, the pocket is fixed to have an area of $1.9\%$ of the Brillouin zone, and the relaxation time is $\tau=0.2 ps$. Only when the mass enhancement is minimal is the Hall coefficient negative at all fields. For large mass enhancement a sign change in the Hall coefficient occurs at a high value of the magnetic field, for instance at $\approx170\ T$ for the case $m_c=8.3 m_e$ shown dashed in blue. When the mass enhancement is substantial, there is also a regime of large $B$-linear magnetoresistance, as described in section \ref{sec:unc}.}
\label{fig:B_dep}
\end{figure}

In the weak field regime, a quasiparticle can only travel a small fraction of a Fermi surface segment before decaying, so \eqref{eq:Chambers0T} can be rewritten as a single integral over the Fermi surface. We will be concerned with the Hall conductivity, which in this limit is conveniently expressed using Ong's ``geometric" formula\cite{Ong1991}:
\begin{equation}
\sigma_{xy}=\frac{2e^3B}{h^2}\cdot A_l,
\end{equation}
where $A_l$ is the signed area swept out by the mean free path vector $\mathbf{l}\equiv\mathbf{v_F}\tau$ as the Fermi surface is traversed. For our model, the mean free path sweeps out one circle of radius $v_l\tau$, (from the light segments) and two circles of radius $v_h\tau$ from the heavy segments, and these contributions have opposite sign:
\begin{align}
\sigma_{xy}=&\frac{2e^3B}{h^2}\left(\pi (v_l\tau)^2-2\pi(v_h\tau)^2\right)\nn\\
=&\frac{2\pi e^3\tau^2 B}{h^2}\left(v_l^2-2v_h^2\right)
\end{align}
Far from the QCP, $v_h\approx v_l$ and the Hall conductivity is negative, as expected for an electron-like pocket. As the QCP is approached, $v_h$ is reduced, eventually yielding a positive Hall conductivity despite the fixed electron-like topology of the pocket.

\subsection{Strong field: $\omega_l\tau,\omega_h\tau\gg 1$}
\begin{figure}
\includegraphics[clip=true, trim=70 230 90 250,width=0.95\columnwidth]{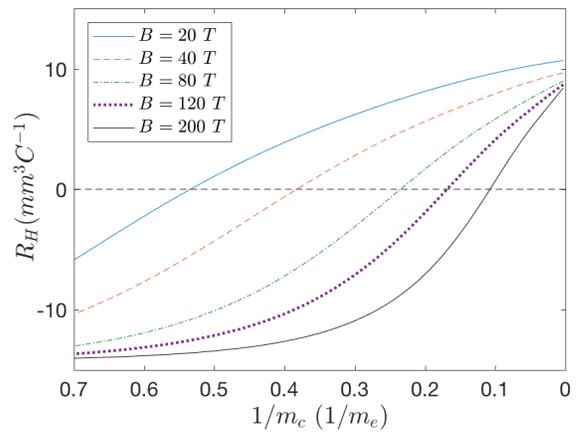}
\caption{The dependence of the Hall coefficient on approach to a QCP at which $m_h$ diverges, with $m_l=m_e$ held fixed. Other parameters are as in Fig. \ref{fig:B_dep}. Data are shown at various values of the magnetic field, showing that the sign change always precedes the QCP.}
\label{fig:m_dep}
\end{figure}

In the strong field regime, a quasiparticle executes numerous cyclotron orbits before decaying, and an expansion of the conductivity in powers of $1/B$ can be obtained by Taylor expanding the exponential of Eq.\eqref{eq:Chambers0T}. Standard manipulations relate the the conductivity to the area of the Fermi surface as

\begin{align}
\sigma_{ij}=-\frac{eA}{2\pi^2 B}\epsilon_{ij}+{\cal O}(1/B^2)
\end{align}
Here $\epsilon$ is the Levi-Civita symbol and $A$ is the signed area of the pocket(s), defined so that an electron (hole) pocket has positive (negative) area. The Hall conductivity goes like $1/B$ at large field, while the closure of the Fermi surface prohibits a $1/B$ term in the longitudinal conductivity, which goes as $1/B^2$. Accordingly, the Hall coefficient obtains its familiar classical expression
\begin{equation}
R_H=\frac{\rho_{yx}}{B}\approx\frac{1}{B\sigma_{xy}}\approx -\frac{1}{ne},
\end{equation}
where $n=2A/(2\pi)^2$ is the signed number density of carriers. Meanwhile, the longitudinal resistivity saturates: $\rho_{xx}\approx \sigma_{xx}/\sigma_{xy}^2\sim B^0$
\section{``Fake" high field regime: $\omega_h\tau\ll 1\ll\omega_l\tau$}
\label{sec:unc}
In a typical metal, the only asymptotic regimes of magnetotransport are the weak- and strong-field regimes described above, with a crossover between them when the cyclotron period is of order $\tau$. In the presence of large mass anisotropy, an additional asymptotic regime exists, in which a quasiparticle does not complete a full cyclotron orbit before decaying, but rapidly traverses parts of the Fermi surface of low effective mass. In our model, transport coefficients in this regime can be expressed as a double expansion in $(\omega_l\tau)^{-1}$ and in $\omega_h\tau$. The zeroth order term in this expansion involves only the geometric properties of the light segments, so that $k_l$ is the only model parameter that enters.

To obtain this zeroth order term, we set the exponential damping factor in Eq. \eqref{eq:Chambers0T} to unity when $\mathbf{q}(t)$ lies in a light segment, and zero when it lies in a heavy segment. We rewrite the integration measure $dt = dq/\dot{|\mathbf{q}|}=\hbar dq/(e|\mathbf{v}(\mathbf{q})|B)$ and write the conductivity as a sum over light segments:
%\begin{widetext}
\begin{align}
\sigma_{ij}=\frac{e}{2\pi^2B}\sum_{\text{light segments } \alpha}&\int_{\mathbf{k}^{\alpha}_0}^{\mathbf{k}^{\alpha}_1} \frac{dk}{\left|\mathbf{v}(\mathbf{k})\right|} v_j(\mathbf{k})\nn\\
\cdot&\int_{\mathbf{k}}^{\mathbf{k}^{\alpha}_1} \frac{dq}{\left|\mathbf{v}(\mathbf{q})\right|} v_i(\mathbf{q}),
\end{align}
where the light segment $\alpha$ begins at $\mathbf{k}^{\alpha}_0$ and ends at $\mathbf{k}^{\alpha}_1$. Evidently, both the longitudinal and Hall conductivity are proportional to $1 /B$ at this level of approximation. For the model in question, the Hall coefficient and longitudinal resistivity are
\begin{align}
R_H=&\frac{2\pi^2}{ek_l^2}\left(\frac{\pi-2}{(\pi-2)^2+4}\right)\left[1+{\cal O}\left(\omega_h\tau, (\omega_l\tau)^{-1}\right)\right]\\
\rho_{xx}=&\frac{4\pi^2 B}{ek_l^2}\left(\frac{1}{(\pi-2)^2+4}\right)\left[1+{\cal O}\left(\omega_h\tau, (\omega_l\tau)^{-1}\right)\right]
\label{eq:fake}
\end{align}
The Hall coefficient remains positive, and is about $0.68$ times its weak field value, while the longitudinal resistivity exhibits an unusual $B$-linear magnetoresistance. The latter phenomenon has been shown to arise near density-wave QCPs when mass enhancement effects are neglected\cite{Fenton2005,Koshelev2013}, due to the sharp curvature of the Fermi surface near the hot spots. $B$-linear magnetoresistance arises in our context in a regime of higher magnetic field, and from a mechanism in which Fermi surface curvature plays no direct role\footnote{Furthermore, B-linear magnetoresistance due to curvature is absent if the Fermi velocity at the turning points is sufficiently small, as is the case in our study}. Unlike the curvature effect, it is also apparent for current flowing perpendicular to the two-dimensional plane treated in this work.

\section{Discussion}
A wealth of experimental evidence points to QCP near $p=0.19$ in YBCO\cite{TallonLoram}, one which is likely relevant for superconductivity\cite{ramshaw2014} and for the strange metal regime that prevails at temperatures above $T_c$. At high magnetic fields and lower dopings, $0.08<p<0.16$, a metallic state with a small electron pocket\cite{Sebastian2015} and CDW order is well established. The high field measurements of Ref. \cite{Badoux2016} cover the doping region between this metal and the QCP, and show strong doping dependence of the Hall number, as well as a change of its sign relative to that at $p<0.16$.

The strong doping dependence of the Hall number is striking, but it does not obviously constrain theories of the QCP. A cusp-like singularity in the Hall number is predicted, within Boltzmann theory, in theories of a variety of order parameter transitions, including d-density wave\cite{Chakravarty2002}, spin-density wave\cite{Coleman2001,Storey2016,Eberlein2016}, and nematicity\cite{Maharaj2016}, among others. A transition between a Fermi liquid and fractionalized Fermi liquid (FL$^*$) state would be expected to feature a discontinuous jump in the Hall number, but this would be inevitably rounded by finite temperature.

The sign change of the Hall number has been interpreted by the authors of \cite{Badoux2016} to rule out CDW order at dopings above $p=0.16$. However, our calculations show that a sign change of the Hall number {\it precedes} the loss of CDW order under reasonable assumptions about Fermi surface anisotropy. Therefore, it is premature to rule out CDW order in the region $0.16<p<0.19$. High field measurements directly sensitive to charge order would clearly be useful to better understand the phenomenology in this doping regime.

\begin{figure}
\includegraphics[clip=true, trim=50 200 70 230,width=0.9\columnwidth]{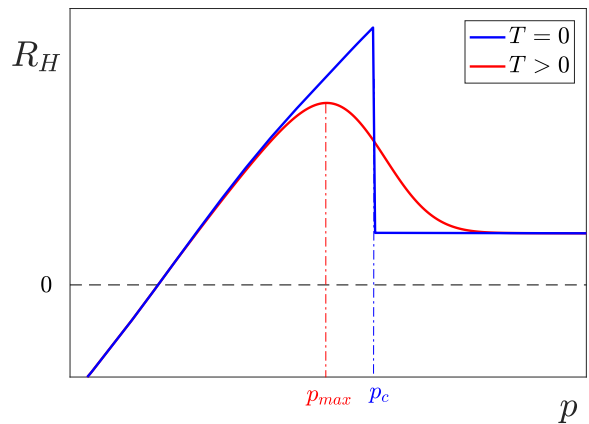}
\caption{A schematic illustration of the effects of finite temperature on the Hall coefficient under the scenario explored in this paper. A small electron pocket persists until a critical doping $p_c$, with the Hall coefficient changing sign well below $p_c$ due to mass renormalization and curvature effects. At zero temperature, there is a sharp jump of the Hall coefficient at $p_c$, where the electron pocket transforms into the large Fermi surface. At finite temperature, this jump is rounded, leaving a maximum in the Hall coefficient at a doping $p_{max}<p_c$. }
\label{fig:schematic}
\end{figure}
{Though our calculations rationalize the sign change without postulating an additional phase in the region $0.16<p<0.19$, the rapid rise of the Hall number with doping in this region requires a different explanation. It is natural to postulate that this is tied to thermal crossover physics around the QCP near $p_c= 0.19$, but we do not attempt an analysis of that QCP here. However it is plausible that $R_H$ has a sharp drop as the doping is increased through $p_c$ at very low $T$.  At higher $T$, this will then lead to a rounded peak in $R_H$, which when combined with our calculations at lower $p$, can lead to the observed behavior. A schematic plot of the Hall coefficient versus doping in this scenario is shown in Fig. \ref{fig:schematic}.}

The results of \cite{Badoux2016} underscore the centrality of a QCP near $p=0.19$  to the phenomenology of YBCO, but evidently offer minimal phenomenological constraints on theories of that QCP. That said, such phenomenological constraints do exist in the literature.  {   Some of these have been discussed in \cite{Senthil2014}, where various general possibilities for the $T = 0$ evolution from the overdoped to the underdoped metal were described.} The absence of an elastic peak near $(\pi,\pi)$ in neutron scattering renders a spin-density wave or d-density wave transition less likely, though neutron measurements in high magnetic field would be necessary to definitively rule out such proposals. { Also, the presence of anti-nodal electron pockets just below a continuous density wave transition precludes the identification of $p= 0.19$ with the opening of the pseudogap, which is conventionally understood to be an anti-nodal phenomenon.  %The anti-nodal electron pockets will have difficulty with the idea that $p = 0.19$ is associated with the opening of the pseudogap at the anti-nodes as suggested by ARPES and other probes (again though we know this only at zero field).
At lower dopings, close to $p = 0.10$, the presence of both nodal and anti-nodal pockets would likely result in electronic specific heat in excess of that measured\cite{Riggs2011}. }

These  problems are mitigated, though not solved entirely, by the scenario of an FL to FL$^*$ transition\footnote{{ Case B of Ref. \onlinecite{Senthil2014}. The comments below also pertain to Cases C and D of that paper.}}. Such a transition can be accompanied by a discontinuous jump in the size of the Fermi surface even when the transition is continuous, yielding an anti-nodal gap, and lower specific heat than the continuous onset of an order parameter. However, it is not known whether the FL to FL$^*$ transition can be continuous, {and novel experiments will be necessary to establish fractionalization in this doping regime.} %the exotic physics of the FL$^*$ state should entail an enhanced burden of proof which has not yet been met by experiments.
Evidently, the results of \cite{Badoux2016} and the analysis pursued in this paper leave considerable uncertainty in the phenomenological constraints on a QCP in the near-optimally doped cuprates.

Moving beyond the cuprates, our calculations point out a previously unappreciated regime of magnetotransport, with a novel orbital mechanism for $B$-linear magnetoresistance. That phenomenon is observed in a wide variety of correlated metals \cite{Jin1999,Lippman2008,Kikugawa2010,Hayes2016}, but often inadequately understood. Our results motivate a more detailed experimental account of the anisotropy of mass enhancement in correlated metals.

\label{sec:disc}

\begin{acknowledgments}
We acknowledge fruitful discussions with Steven Kivelson, Francis Lalibert\'e, Subir Sachdev, and Louis Taillefer. S.~L. was supported by a Gordon and Betty Moore Postdoctoral Fellowship. L.~Z. and T.~S. were supported by DOE grant  DE-SC0008739. T.~S was also partially supported by a Simons Investigator Award from the Simons Foundation.
\end{acknowledgments}

\begin{appendix}

\section{Review of the conventional regimes}

In this appendix, we review the conductivities in the conventional regimes predicted by Chambers' formula in some detail:
\beq \label{eq:Chambersapp}
\begin{split}
\sigma_{ij}=&\frac{e^2}{2\pi^2\hbar}\oint\frac{dk}{|\bf v(\bf k)|}v_j({\bf k})\\
&\qquad\qquad\qquad
\cdot\int_0^\infty dtv_i({\bf q}(t))e^{-\int_0^t\frac{dt'}{\tau({\bf q}(t'))}}
\end{split}
\eeq
As in the main text, we will focus on the two dimensional case. To define the weak and strong field regimes, we introduce local cyclotron frequency $\omega({\bf q})\equiv eB/m({\bf q})$, where ${\bf q}$ lies on the Fermi surface and $m({\bf q})$ is the effective mass at that point. For generality, we also introduce $\tau({\bf q})$, which is the local relaxation time of electrons at point ${\bf q}$.

We will see that in the strong field regime, the Hall coefficient is indeed related to the carrier type and carrier density. In the weak field regime, although they are not related in general, in the special case of isotropic Fermi surfaces, the Hall coefficient happens to be related to the carrier type and carrier density in the same way as in the strong field regime.

\subsection{Weak field regime}

We start by reviewing the weak field regime, where $\omega({\bf q})\tau({\bf q})\ll 1$ for all ${\bf q}$ on the Fermi surface. In this regime, (\ref{eq:Chambersapp}) can be expanded in powers of $B$:
\beq
\begin{split}
v_i({\bf q}(t))&=v_i({\bf k})-\nabla_{\bf k}v_i({\bf k})\cdot\frac{e{\bf v}({\bf k})\times{\bf B}}{\hbar}t+\mc{O}(B^2)\\
\tau({\bf q}(t))&=\tau({\bf k})-\nabla_{\bf k}\tau({\rm k})\cdot\frac{e{\bf v}({\bf k})\times{\bf B}}{\hbar}t+\mc{O}(B^2)
\end{split}
\eeq
which yields
\beq
\begin{split}
&\int_0^\infty dtv_i({\bf q}(t))e^{-\int_0^t\frac{dt'}{\tau({\bf q}(t'))}}\\
=&v_i({\bf k})\tau({\bf k})-\tau({\bf k})\frac{e{\bf v}\times{\bf B}}{\hbar}\nabla_{\bf k}(v_i({\bf k})\tau({\bf k}))+\mc{O}(B^2)
\end{split}
\eeq

So the weak-field limit of (\ref{eq:Chambersapp}) is
\beq
\begin{split}
\sigma_{ij}
=&\frac{e^2}{2\pi^2\hbar}\oint dq\frac{v_j({\bf q})v_i({\bf q})\tau({\bf q})}{|{\bf v}({\bf q})|}\\
&-\frac{e^3B}{2\pi^2\hbar}\oint dql_j({\bf q})\hat t\cdot\nabla_{\bf q}l_i({\bf q})
\end{split}
\eeq
where $\hat t$ is the tangential direction of the Fermi surface at $\bf q$ and $l_i(\bf q)=v_i(\bf q)\tau(\bf q)$.

If $i=j$, the second term vanishes for a closed Fermi surface, so we get
\beq
\sigma_{xx}=\frac{e^2}{2\pi^2\hbar}\oint dq\frac{v_x^2(\bf q)\tau(\bf q)}{|\bf v(\bf q)|}+\mc{O}(B^2)
\eeq
With time reversal symmetry, the first term vanishes if $i\neq j$, and we recover Ong's formula for Hall conductivity
\beq \label{eq:Ong}
\begin{split}
\sigma_{xy}
=&-\frac{e^3B}{2\pi^2\hbar}\oint dql_y(\vec q)\hat t\cdot\nabla_{\vec q}l_x(\vec q)\\
=&\frac{e^3}{2\pi^2\hbar}\vec B\cdot\oint d\vec l\times\vec l/2=2\frac{e^2}{h}\frac{\phi}{\phi_0}
\end{split}
\eeq
where $\phi=\vec B\cdot\oint d\vec l\times\vec l/2$ and $\phi_0=h/e$.

In general in this regime the Hall number is not related to the carrier type and carrier density, unless the Fermi surface happens to be isotropic. To see this, notice for circular isotropic Fermi surfaces, we have
\beq
\sigma_{xx}=\frac{e^2k_Fv_F\tau}{2\pi\hbar}
\eeq
where $k_F$ and $v_F$ are the Fermi momentum and Fermi velocity, respectively, and $\tau$ is the scattering rate on the Fermi surface. Assuming a quadratic band with effective mass $m$, we have $k_F^2=2\pi n$ and $\hbar k_F=mv_F$, where $n$ is the carrier density. Then we get
\beq
\sigma_{xx}=\frac{ne^2\tau}{m}
\eeq
which is Drude's formula for conductivity for $\sigma_{xx}$.

As for $\sigma_{xy}$, in this case $\phi=-BA\frac{\hbar^2\tau^2}{m^2}$, where $A$ is the area of the Fermi surface, which is related to the carrier density by Luttinger's theorem: $A=2\pi^2n$. Plugging these in, we find
\beq
\sigma_{xy}=-\frac{e^3\tau^2Bn}{m^2}
\eeq

So the Hall number is
\beq
R_H=\frac{\rho_{yx}}{B}=\frac{1}{B}\frac{\sigma_{xy}}{\sigma_{xx}^2+\sigma_{xy}^2}
\approx\frac{1}{B}\frac{\sigma_{xy}}{\sigma_{xx}^2}\approx-\frac{1}{ne}
\eeq
Therefore, in the case of isotropic Fermi surface the Hall number can still indicate the carrier type and carrier density even in the weak field regime.

\subsection{Strong field regime}

Now we turn to the strong field regime, where $\omega(\bf q)\tau(\bf q)\gg 1$ for all $\bf q$ on the Fermi surface. To study the conductivities in the strong field regime, we first write (\ref{eq:Chambersapp}) in another form by using that $\frac{\hbar d\bf q}{dt}=-e{\bf v}\times{\bf B}$:
\beq
\begin{split}
\sigma_{ij}
=&\frac{e^2}{2\pi^2\hbar}\oint\frac{dk}{|{\bf v}({\bf k})|}v_j({\bf k})\\
&\qquad
\cdot\int\frac{\hbar dq}{eB|{\bf v}({\bf q})|}v_i({\bf q})e^{-\int_0^t\frac{dt'}{\tau({\bf q(t')})}}
\end{split}
\eeq
The second integral over momentum goes over the Fermi surface repeatedly. Therefore, we can limit the second integral to be the first complete cyclotron motion and represent the following cyclotron motion as a geometric series:
\beq
\begin{split}
\sigma_{ij}=&\frac{e}{2\pi^2 B}\oint\frac{dk}{|{\bf v}({\bf k})|}v_j({\bf k})\\
&\
\cdot\oint\frac{dq}{|{\bf v}({\bf q})|}{v_i}({\bf q})e^{-\int_0^t\frac{dt'}{\tau({\bf q(t')})}}\cdot\left[\frac{1}{1-e^{-T/\bar\tau}}\right]
\end{split}
\eeq
where $T$ is the cyclotron period and $\bar\tau$ is defined such that $T/\bar\tau=\int_0^Tdt'/\tau({\rm q}(t'))$.

In the strong field regime, the electrons travel so fast that $T\sim 1/B\ll\bar\tau$, we can expand the above in terms of $1/B$:
\beq
\begin{split}
\sigma_{ij}
\approx&\frac{e}{2\pi^2B}\oint\frac{dk}{|{\bf v}({\bf k})|}v_j({\bf k})\\
&\quad
\cdot\oint\frac{dq}{|\bf v({\bf q})|}v_i({\bf q})\left(1-\int_0^{t({\bf q})}\frac{dt'}{\tau({\bf q(t')})}\right)\frac{\bar\tau}{T}\\
=&\frac{e}{2\pi^2B}\oint \epsilon_{jl}dk_l\oint\epsilon_{im}dq_m \left(1-\int_0^{t({\bf q})}\frac{dt'}{\tau({\bf q(t')})}\right)\frac{\bar\tau}{T}
\end{split}
\eeq
The integral involving the first term vanishes for closed Fermi surfaces, because the result of the integral over ${\bf q}$ will be independent of ${\bf k}$, and $\oint dk_l$ vanishes for a closed Fermi surface. For the second term, we integrate by parts
\beq
\begin{split}
&-\oint\epsilon_{im}dq_m\int_0^{t({\bf q})}\frac{dt'}{\tau({\bf q}(t'))}\frac{\bar\tau}{T}\\
&\qquad\qquad
=-\epsilon_{im}k_m+\int_0^T \frac{dt}{\tau({\bf q}(t))}\frac{\epsilon_{im}q_m\bar\tau}{T}
\end{split}
\eeq
Again, the second term does not depend on ${\bf k}$ so it does not contribute after the integration over ${\bf k}$. So we get
\beq
\sigma_{ij}=-\frac{e}{2\pi^2B}\epsilon_{jl}\epsilon_{im}\oint dk_lk_m=-\frac{e}{2\pi^2B}\epsilon_{ij}A
\eeq
This result tells us that for a closed Fermi surface in the strong field regime, the leading nonzero contribution of the off-diagonal elements of the conductivity tensor is at the order $1/B$, but leading contribution to the diagonal elements will be at order $1/B^2$ or higher.

In this regime, the Hall number is given by
\beq
R_H=\frac{\rho_{yx}}{B}\approx\frac{1}{B\sigma_{xy}}=-\frac{1}{ne}
\eeq
where Luttinger's theorem, $A=2\pi^2n$, is applied in the last step. So in the strong field regime the Hall number is always related to the carrier type and carrier density, as long as the Fermi surface is closed.

%\section{The unusual regime}
%
%The unusual regime is where there is a significant portion of Fermi surface that satisfies $\omega({\bf q})\tau({\bf q})\gg 1$ while cyclotron period is still much larger than the averaged value of scattering rate. In this regime, we can divide the entire Fermi surface into fast segments, where electrons can traverse quickly so that the exponential in (\ref{eq:Chambersapp}) is approximately one, and slow segments, where electrons are locally in the weak field regime.
%
%In this regime, assume a sharp separation between fast and slow segments, then the leading contribution is
%\beq
%\sigma_{ij}=\frac{e}{2\pi^2B}\sum_{\rm fast\ segments}\int_{\rm fast\ segments}dk\frac{v_j({\bf k})}{|{\bf v}({\bf k})|}\int_{\rm q\ after\ k} dq\frac{v_i({\bf q})}{|{\bf v}({\bf q})|}
%\eeq

\section{Corrections to the conductivities at the order of $(\omega_l\tau)^{-1}$ and $\omega_h\tau$}

In the main text the conductivities of the specific model we consider have been calculated to the leading order in $(\omega_l\tau)^{-1}$ and $\omega_h\tau$. In this appendix we give the leading corrections to those results.

The leading corrections come from two sources. First, it comes from the higher order terms in the expansion of the exponential, and this will give a correction at the order of $(\omega_l\tau)^{-1}$. Second, it comes from the leakage among fast segments and slow segments, and this will give a correction at the order of $\omega_h\tau$.

To this order, we get conductivities
\beq
\begin{split}
\sigma_{xx}=&\frac{2e}{B}\Bigg[\frac{k_l^2}{2\pi^2}+\frac{k_l^2}{\pi^2} \left(\frac{\pi}{4}-1\right)\frac{1}{\omega_l\tau}\\
&\qquad\qquad
+\left(\frac{k_lk_h}{\pi^2}+\frac{k_h^2}{2\pi}\right)\omega_h\tau\Bigg]
\end{split}
\eeq
\beq
\begin{split}
\sigma_{xy}=&\frac{2e}{B}\Bigg[\frac{k_l^2}{2\pi^2} \left(\frac{\pi}{2}-1\right)-\frac{k_l^2}{2\pi^2}\frac{4-\pi}{2} \frac{1}{\omega_l\tau}\\
&\qquad\qquad\qquad\qquad\qquad
+\frac{k_lk_h}{\pi^2}\omega_h\tau\Bigg]
\end{split}
\eeq
which yields resistivities
\beq
\begin{split}
\rho_{xx}
=&\frac{\sigma_{xx}}{\sigma_{xx}^2+\sigma_{xy}^2}\\
=&\frac{B}{2ek_l^2}\left(C_1+C_2\frac{1}{\omega_l\tau}+C_3\omega_h\tau\right)\\
\rho_{yx}
=&\frac{\sigma_{xy}}{\sigma_{xx}^2+\sigma_{xy}^2}\\
=&\frac{B}{2ek_l^2}\left(D_1+D_2\frac{1}{\omega_l\tau}+D_3\omega_h\tau\right)
\end{split}
\eeq
with
\beq
\begin{split}
C_1&=\frac{8\pi^2}{(\pi-2)^2+4}\\
C_2&=\frac{4(4-\pi)\pi^2(\pi^2-8)}{((\pi-2)^2+4)^2}\\
C_3&=\frac{8\pi^2\left((\pi-4)\pi^2\alpha^2+2(\pi(\pi-8)+8)\alpha\right)} {((\pi-2)^2+4)^2}
\end{split}
\eeq
and
\beq
\begin{split}
D_1&=\frac{4(\pi-2)\pi^2}{(\pi-2)^2+4}\\
D_2&=\frac{4(4-\pi)\pi^2(\pi^2-8)}{((\pi-2)^2+4)^2}\\
D_3&=\frac{-16\pi^2(2\pi(\pi-2)\alpha^2+(\pi^2-8)\alpha)}{((\pi-2)^2+4)^2}
\end{split}
\eeq
where $\alpha=k_h/k_l$.

\end{appendix}
\clearpage

\end{document}